\documentclass[runningheads]{llncs}

\usepackage[T1]{fontenc}
\usepackage{xspace}

\usepackage{threeparttable}
\usepackage{booktabs}

\usepackage[table]{xcolor}

\usepackage{tikz}
\usepackage{pgfplots}
\usepackage{bbding}
\pgfplotsset{width=8cm,compat=1.9}

\usepackage{amssymb}

\usepackage{float}
\usepackage{graphicx}

\usepackage{multirow}
\usepackage{multicol}
\usepackage{booktabs}
\usepackage{makecell}
\usepackage{tabularx}
\usepackage{longtable}
\usepackage{threeparttable}
\usepackage{wrapfig}
\usepackage{ulem}
\usepackage{colortbl}

\usepackage{csquotes}

\usepackage{soul}

\usepackage{hyperref}
\hypersetup{hidelinks}
\usepackage[ruled,vlined]{algorithm2e}

\usepackage{enumitem}

\usepackage{pifont}

\usepackage{url}
\usepackage{diagbox}
\usepackage{adjustbox}
\usepackage{breqn}

\usepackage{arydshln}
\usepackage{tcolorbox}

\newcommand{\tool}{{UniAE-MoE}\xspace}
\newcommand{\circled}[1]{%
  \begin{tikzpicture}[baseline=(char.base)]
    \node[shape=circle, draw, inner sep=0pt] (char) {#1};
  \end{tikzpicture}%
}

\usepackage{comment}

\begin{document}

%==================================================================================
% Title
\title{\tool: A Unified Audio Encoder via Mixture of Experts}
\titlerunning{UniAE-MoE: A Unified Audio Encoder via Mixture of Experts}

%==================================================================================
% Authors (LLNCS format)
\author{Shengbo Cai\textsuperscript{*} \and
Zhisheng Zhang\textsuperscript{*} \and
Zichao Nie \and
Jing Peng \and
Jingran Xie \and
Zhiyong Wu\Envelope}
\authorrunning{S. Cai et al.}

%==================================================================================
% Affiliation and emails
\institute{Tsinghua University, China\\
\email{csb25@mails.tsinghua.edu.cn, zywu@sz.tsinghua.edu.cn}\\
\textsuperscript{*}Equal contribution.}

\maketitle

\begin{abstract}
    Large Audio Language Models (LALMs) rely on effective audio encoders for multi-task performance. We introduce \textbf{\tool}, a unified audio encoder designed to model cross-domain audio representations and achieve outstanding downstream understanding performance via a Mixture-of-Experts (MoE) architecture.
    Specifically, we explore mainstream audio encoders and integrate those from Qwen2-Audio and Audio-Flamingo 3, which demonstrate superior downstream capabilities. To facilitate effective model fusion, we improve our encoder using SwiGLU with shared experts to decouple encoder networks, and we further introduce a two-stage instruction-tuning strategy to better adapt the model to diverse downstream tasks. Moreover, we propose the task-specific data scaling (TSDS) technique to enhance \tool's understanding capabilities. 
    On the XARES-LLM benchmark, \tool attains a score of \textbf{0.802}, achieving state-of-the-art performance. It also delivers top-tier performance in the official Interspeech 2026 Audio Encoder Capability Challenge, further demonstrating robust generalization across diverse audio tasks. Together, these results validate the effectiveness of \tool for unified audio understanding across speech, music, and general audio domains.

\end{abstract}

\section{Introduction}

Large Audio Language Models are shifting from simple audio-to-text conversion toward ``General Audio Intelligence''. By integrating powerful audio encoders with Large Language Models (LLMs)~\cite{AudioGPT}, recent models, \textit{e.g.}, AudioPaLM~\cite{AudioPaLM} and Kimi-Audio~\cite{kimi-audio}, unify speech, general audio, and music perception. More recently, Qwen2-Audio~\cite{Qwen2} leverages natural language prompts to enhance instruction-following across diverse benchmarks. These advancements underscore the necessity of high-quality audio encoders that provide generalized representations for a broad spectrum of audio tasks.

Despite these advancements, developing a unified audio encoder remains challenging. Traditional pipelines typically separate audio classification from comprehension and generation, often relying on task-specific heads that limit architectural flexibility and impede knowledge transfer across domains. Meanwhile, current LALMs commonly employ monolithic encoders or dense shared adapters~\cite{kimi-audio,Flamingo,whisper}, forcing a single parameter space to capture heterogeneous acoustic characteristics ranging from fine-grained temporal cues to high-level semantics. General-purpose representation models such as AudioMAE~\cite{AudioMAE} and BEATs~\cite{BEATs}, although effective for perception and classification, are not inherently aligned with generative audio-to-language inference and instruction-following objectives. Integrating multiple encoders, as explored by SALMONN~\cite{SALMONN} and WavLLM~\cite{wavllm}, provides access to complementary representations, but straightforward fusion can introduce feature redundancy and performance trade-offs, particularly in speech recognition. Recent advances in MoE-based adapters~\cite{moeadapter,promptmoe} indicate that sparsely routing inputs to specialized experts can decouple audio attributes or sub-domains while retaining shared knowledge, offering a promising direction for robust multi-task audio modeling.

In this work, we introduce \textbf{UniAE-MoE}, a novel unified audio encoder designed to explicitly decouple the often conflicting demands of cross-domain audio using Mixture of Experts. By leveraging instructionalized unified training objectives, we converge multi-task learning into a cohesive generative optimization framework. Concretely, we employ two encoders with top-performance, Qwen2-Audio and Audio-Flamingo 3, and exploit the MoE architecture to decouple the parameter space, allowing experts to specialize in acoustic modalities. To facilitate the extraction of high-quality, task-tailored semantic features from raw audio, we unify 20 downstream tasks into an ``instruction-response'' text sequence generation paradigm and devise a two-stage training strategy that advances from semantic alignment to multi-task instruction fine-tuning. Considering the unbalance of each task, we propose the task-specific data scaling method for better understanding performance. Our methodology achieves robust generalization across speech, music, and ambient sounds, delivering excellent performance in both classification and generation tasks simultaneously. We release our code\footnote{\url{https://github.com/Syclus123/UniAE-MoE}} for future comparison. 
Our contributions are as follows:

\begin{itemize}[itemsep=0pt,topsep=0pt,parsep=0pt,leftmargin=10pt]
    \item We propose \textbf{UniAE-MoE}, a unified audio encoder that integrates dual high-performance backbones (Qwen2-Audio and Audio-Flamingo 3) via a SwiGLU-based MoE fusion module with a shared expert, enabling the effective decoupling and fusion of cross-domain audio representations.
    \item We establish a unified generative paradigm that reformulates 20 diverse downstream tasks into an instruction-following framework, complemented by a \textbf{two-stage instruction-tuning strategy} that facilitates a seamless transition from modality alignment to robust multi-task generalization.
    \item We introduce \textbf{Task-Specific Data Scaling} to mitigate cross-task data imbalance. \tool achieves state-of-the-art performance on the XARES-LLM benchmark with a high score of \textbf{0.802}, and further demonstrates strong generalization in Interspeech 2026 Audio Encoder Capability Challenge, attaining \textbf{top-tier} performance across both classification and understanding tasks.
\end{itemize}

\begin{figure}[!ht]
  \centering
  \IfFileExists{Figure/UniAE-MoE.png}{%
    \includegraphics[width=\linewidth]{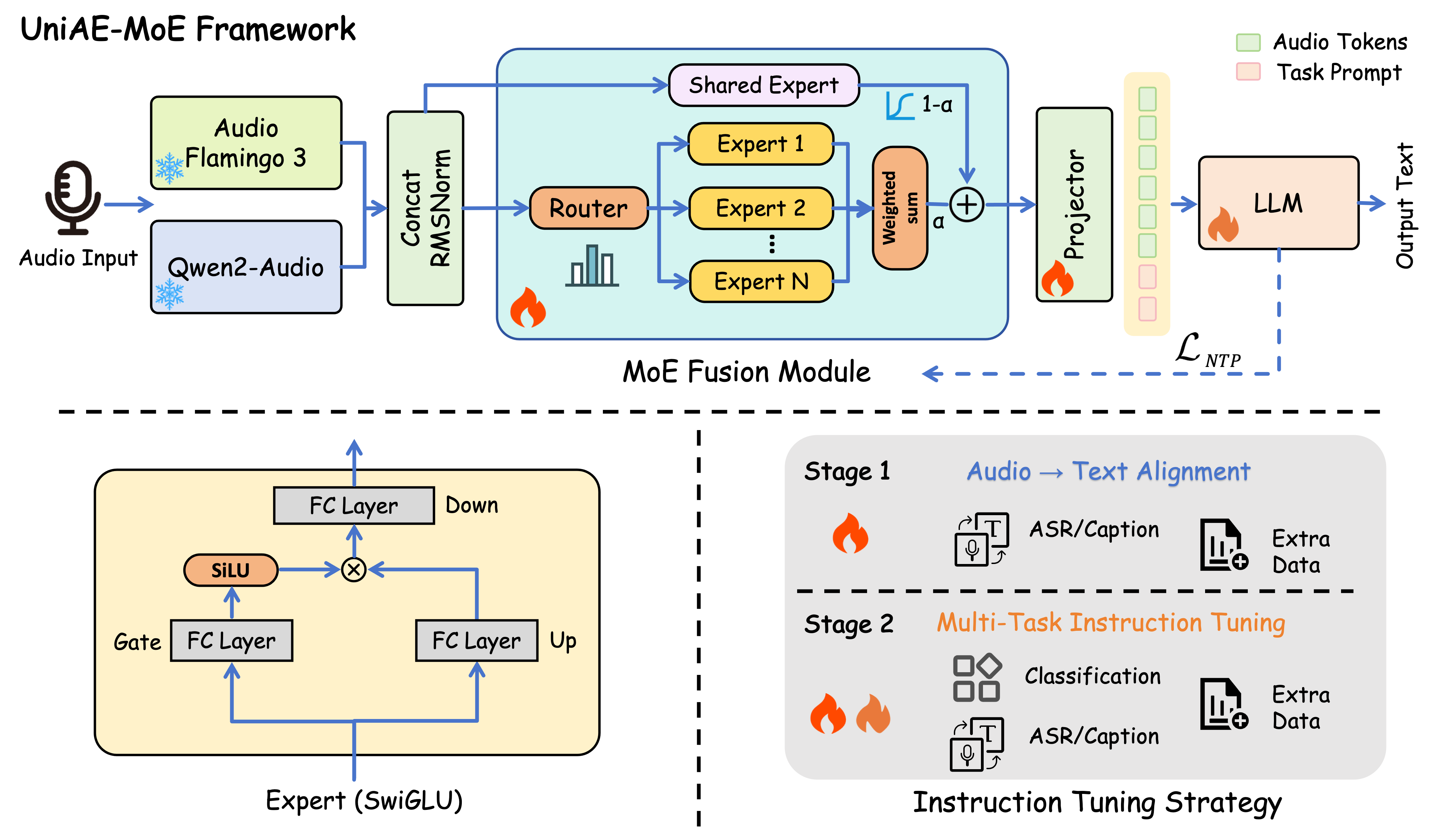}%
  }{%
    \fbox{\begin{minipage}[c][0.26\textheight][c]{0.96\linewidth}\centering\itshape
      [Figure placeholder] Copy \texttt{UniAE-MoE.png} into the \texttt{Figure/} folder to display the framework overview.\end{minipage}}%
  }
  \caption{The UniAE-MoE framework comprises the overall architecture (top), the SwiGLU-based expert network (bottom left), and the two-stage instruction-tuning strategy (bottom right).}
  \label{fig:UniAE-MoE}
  % \vspace{-0.2cm}
\end{figure}

\section{Methodology}

\subsection{High-Performance Dual-Model Backbone}
Using Dasheng-base as a baseline, we benchmark representative encoders, including the Qwen series, Audio-Flamingo, and Whisper. As shown in Table~\ref{tab:encoders}, Audio-Flamingo 3 and Qwen2-Audio achieve the two highest overall scores of 0.767 and 0.737, respectively.

Despite its strong performance, Qwen2-Audio's pre-training distribution places greater emphasis on speech and music, with comparatively limited coverage of general acoustic events. Audio-Flamingo 3, in contrast, exhibits more balanced capabilities for long-form audio and environmental sounds, complementing Qwen2-Audio in general audio modeling. We therefore combine these encoders into a dual-model backbone and investigate their fusion for robust, fine-grained unified audio modeling, with the corresponding analyses presented in Section~\ref{sec:encoders}.

\begin{table}[!ht]
    \caption{Comparison of different audio encoders on the XARES-LLM benchmark}
    \centering
    \label{tab:main_results}
    \setlength{\tabcolsep}{0.5pt}
    \resizebox{\textwidth}{!}{%
    \begin{threeparttable}
    \begin{tabular}{ccccccccccc}
    \toprule
    \multirow{2}{*}{\textbf{Num \#}} 
    & \multirow{2}{*}{\textbf{Task}} 
    & \multicolumn{2}{c}{\textbf{Baseline}}
    & \multicolumn{6}{c}{\textbf{Model Fusion (Concat)}}
    & \multirow{2}{*}{\textbf{\makecell[c]{\tool \\ (Ours)}}}  \\
    % \midrule
    \cmidrule(r){3-4} \cmidrule(lr){5-10}
    & & Dasheng-Base & Whisper-Base 
    & (Q2A, DS) & (Q2A, KA) & (Q2A, AF3)
    & (Q2.5O, DS) & (AF3, KA) & (Q2A, AF3, DS)
    & \\
    \midrule
    \rowcolor{gray!10} \multicolumn{11}{c}{\textit{Task 1 Performance}} \\
    \midrule
    \circled{1} & ASVSpoof 2015 
        & 0.937 & 0.943 
        & 0.980 & 0.978 & 0.985 & 0.986 & \textbf{0.995} & 0.994
        & 0.982 \\
    \circled{2} & CREMA-D 
        & 0.621 & 0.516 
        & 0.732 & 0.795 & 0.855 & 0.626 & \textbf{0.858} & 0.808
        & 0.847 \\
    \circled{3} & ESC-50 
        & 0.755 & 0.635 
        & 0.900 & 0.898 & \textbf{0.932} & 0.865 & 0.922 & 0.915
        & 0.882 \\
    \circled{4} & Fluent Speech Commands 
        & 0.984 & 0.817 
        & 0.983 & 0.992 & \textbf{0.994} & 0.993 & 0.982 & 0.993
        & \textbf{0.994} \\
    \circled{5} & Free Music Archive 
        & 0.429 & 0.579 
        & 0.718 & 0.720 & 0.887 & 0.632 & 0.723 & 0.904
        & \textbf{0.979} \\
    \circled{6} & FSD50K 
        & 0.063 & 0.092 
        & 0.201 & 0.221 & 0.239 & 0.150 & 0.236 & 0.239
        & \textbf{0.251} \\
    \circled{7} & FSDKaggle2018
        & 0.415 & 0.552 
        & 0.770 & 0.798 & 0.852 & 0.727 & 0.832 & 0.851
        & \textbf{0.872} \\
    \circled{8} & GTZAN 
        & 0.323 & 0.697 
        & \textbf{0.939} & 0.909 & 0.869 & 0.859 & 0.899 & 0.889
        & 0.899 \\
    \circled{9} & LibriCount 
        & 0.386 & \textbf{0.409} 
        & 0.108 & 0.150 & 0.190 & 0.115 & 0.116 & 0.161
        & 0.381 \\
    \circled{10} & NSynth 
        & 0.675 & 0.638 
        & 0.753 & 0.681 & 0.784 & 0.710 & 0.784 & \textbf{0.788}
        & 0.782 \\
    \circled{11} & Speech Commands V1 
        & 0.655 & 0.694 
        & 0.929 & 0.927 & 0.936 & 0.959 & 0.928 & 0.931
        & \textbf{0.968} \\
    \circled{12} & UrbanSound8K 
        & 0.829 & 0.737 
        & 0.846 & 0.828 & \textbf{0.898} & 0.864 & 0.897 & 0.883
        & 0.888 \\
    \circled{13} & VocalSound 
        & 0.855 & 0.867 
        & 0.941 & 0.938 & \textbf{0.948} & 0.935 & 0.947 & 0.947
        & 0.941 \\
    \circled{14} & VoxCeleb1 
        & 0.974 & 0.762 
        & 0.970 & 0.973 & 0.977 & 0.627 & 0.980 & \textbf{0.982}
        & 0.974 \\
    \circled{15} & VoxLingua33 
        & 0.311 & 0.835 
        & 0.876 & \textbf{0.960} & 0.897 & 0.886 & 0.952 & 0.890
        & 0.940 \\ 
    \hdashline
    & \textbf{Task 1} 
        & 0.614 & 0.652 
        & 0.777 & 0.785 & 0.816 & 0.729 & 0.803 & 0.812
        & \textbf{0.839} \\ 
    \midrule
    \rowcolor{gray!10} \multicolumn{11}{c}{\textit{Task 2 Performance}} \\
    \midrule
    \circled{16} & AISHELL-1 
        & 0.018 & 0.361 
        & 0.700 & 0.780 & 0.796 & 0.733 & 0.388 & 0.718
        & \textbf{0.882} \\
    \circled{17} & Clotho 
        & 0.207 & 0.358 
        & 0.455 & 0.458 & \textbf{0.469} & 0.418 & 0.460 & 0.465
        & 0.449 \\
    \circled{18} & LibriSpeech 
        & 0.103 & 0.385 
        & 0.824 & 0.882 & 0.881 & 0.843 & 0.396 & 0.832
        & \textbf{0.914} \\
    \circled{19} & MECAT 
        & 0.600 & 0.624 
        & 0.616 & 0.617 & 0.660 & 0.620 & 0.646 & 0.647
        & \textbf{0.700} \\
    \circled{20} & SongDescriber 
        & 0.410 & 0.448 
        & 0.494 & 0.502 & 0.503 & 0.471 & \textbf{0.525} & 0.494
        & 0.512 \\ 
    \hdashline
    & \textbf{Task 2} 
        & 0.270 & 0.400
        & 0.618 & 0.648 & 0.662 & 0.617 & 0.483 & 0.613
        & \textbf{0.691} \\ 
    \midrule
    \rowcolor{gray!10} \multicolumn{11}{c}{\textit{Overall Performance (Task 1 + Task 2)}} \\
    \midrule
    & \textbf{Overall} 
        & 0.528 & 0.597
        & 0.737 & 0.750 & 0.778 & 0.708 & 0.723 & 0.766
        & \textbf{0.802} \\ 
    \bottomrule
    \end{tabular}%
    \begin{tablenotes}
        \item {\emph{Model Abbreviation.} 
        (1) \textbf{Q2A}: Qwen2-Audio. 
        (2) \textbf{DS}: Dasheng-1.2B.
        (3) \textbf{KA}: Kimi-Audio.
        (4) \textbf{AF3}: Audio-Flamingo 3.
        (5) \textbf{Q2.5O}: Qwen2.5-Omni.
        } 
    \end{tablenotes}
    \end{threeparttable}
}
% \vspace{-0.2cm}
\end{table}

\subsection{Universal Audio Representation with MoE Fusion}
Across heterogeneous tasks such as speech recognition, audio classification, and audio captioning, a single encoder often struggles to simultaneously preserve fine-grained acoustic details and high-level semantics. Although Audio-Flamingo 3 and Qwen2-Audio share a Whisper-based backbone, their distinct pre-training objectives and instruction-tuning strategies encourage them to encode complementary acoustic and semantic information. To effectively reconcile these heterogeneous features, we introduce a Mixture-of-Experts (MoE) framework to derive universal audio representations. As illustrated in Fig.~\ref{fig:UniAE-MoE}, the routing mechanism dynamically activates specialized experts according to the input characteristics, allowing different audio attributes to be modeled in separate parameter spaces while retaining task adaptability and parameter efficiency.

Specifically, we project the feature sequences $F_{AF}$ from Audio-Flamingo 3 and $F_{QW}$ from Qwen2-Audio into a unified embedding space. 
Both audio sequences undergo temporal and feature-wise alignment, followed by concatenation and normalization along the feature dimension to obtain the composite feature vector $x$. 
Because the concatenated dual-encoder features remain heterogeneous and audio signals contain substantial time--frequency redundancy, we instantiate each expert with SwiGLU~\cite{swiglu}. Its feature-wise gating and multiplicative interactions selectively preserve informative dimensions while reconciling acoustic and semantic features from the two encoders:

\begin{equation}
    \operatorname{SwiGLU}(x)
    =
    \left[
    \operatorname{SiLU}(xW_{\mathrm{gate}})
    \odot
    (xW_{\mathrm{up}})
    \right]W_{\mathrm{down}},
    \end{equation}
where $W_{gate}$, $W_{up}$, and $W_{down}$ denote the learnable parameter matrices within the expert layer. 
The element-wise multiplication between the two projection branches
introduces multiplicative feature interactions, enabling each expert
to more effectively integrate heterogeneous representations from the
two audio encoders.
Unlike a dense feed-forward layer that applies the same set of
parameters to every token, our MoE module performs token-level
conditional expert aggregation. Specifically, an MLP-based router
computes routing scores for each token $x_t$ over $N$ experts. The
top-$K$ experts are then selected, and their outputs are combined
using normalized routing weights:

\begin{equation}
y_{routed} = \sum_{i=1}^{K} \text{Softmax}(G(x_t))_i \cdot E_i(x_t),
\end{equation}
where $E_i$ denotes the $i$-th selected SwiGLU expert. Activating only a subset of experts encourages domain- and task-specific specialization while improving the computation-to-capacity ratio for efficient inference.

To further enhance modeling robustness in complex audio environments, we incorporate a Shared Expert mechanism. Distinct from the experts dynamically activated via routing, the shared expert $E_{shared}$ processes every token, aiming to capture invariant acoustic features across diverse tasks. 
The final output representation is a dynamic reconciliation of the routed result and the shared expert’s output, modulated by a learnable gating parameter $g_{shared}$:

\begin{equation}
\begin{aligned}
\alpha &= \sigma(g_{\text{shared}}), \\
Y &= \mathrm{RMSNorm}\left(\alpha\, y_{\text{routed}} + (1-\alpha\,) E_{\text{shared}}(x) \right).
\end{aligned}
\end{equation}
By doing so, we not only preserve the complementarity of features across different models but also mitigate the issue of expert collapse through the shared expert, yielding more robust representations for multi-task audio understanding.

\subsection{Multi-task Instruction Tuning}

To support diverse audio tasks within a single architecture, we reformulate 20 downstream tasks under a unified \textit{instruction-following} paradigm. Specifically, both discriminative tasks, whose labels are serialized as textual responses, and open-ended generation tasks are cast as autoregressive conditional sequence modeling problems. This formulation eliminates task-specific prediction heads and enables joint optimization using the standard Next Token Prediction (NTP) objective of the LLM. Under this shared objective, end-to-end training encourages the MoE router to retain semantically informative features across task domains. Formally, given an audio sequence $A$ and a corresponding instruction $I$, the model generates a target response $Y = \{y_1, \dots, y_L\}$ by maximizing its conditional likelihood:
\begin{equation}
P(Y \mid A, I; \theta) = \prod_{i=1}^{L} P(y_i \mid y_{<i}, \mathbf{F}_{moe}, I; \theta),
\end{equation}
where $\mathbf{F}_{moe}$ denotes the universal audio representation produced by MoE fusion. To bridge the modality gap between raw acoustic features and the LLM's embedding space, we employ a two-stage training strategy that provides a stable transition from low-level acoustics to high-level semantic understanding.

\begin{table}[!ht]
    \centering
    \caption{Statistics of our training data}
    \label{tab:data_counts}
    % \usepackage{graphicx}, \usepackage{booktabs},\usepackage{multirow}
    % \footnotesize
  \scriptsize
    \renewcommand{\arraystretch}{1.02}
    \begin{tabular*}{\linewidth}{@{\extracolsep{\fill}}llrc@{}}
      % \toprule[1pt]\midrule[0.3pt]
      \toprule
      \textbf{Task} & \textbf{Dataset} & \textbf{Train \#} & \textbf{New} \\ 
      \midrule
      \multicolumn{4}{c}{\textit{Task 1 for Stage-2 Training}} \\
      \midrule
      Spoofing Detection & ASVSpoof 2015 & 15,720 & -\\
      \multirow{2}{*}{Emotion Recognition} & CREMA-D & 6,000 & - \\
          & \cellcolor{gray!10}IEMOCAP/TESS/RAVDESS & \cellcolor{gray!10}3,552 & \cellcolor{gray!10}\ding{52} \\
      Environment Class & ESC-50 & 1,600 & - \\
      Intent Classification & Fluent Speech Commands & 23,132 & - \\ 
      \multirow{2}{*}{Music Genre Class} & Free Music Archive & 6,400 & - \\
          & \cellcolor{gray!10}Free Music Archive (medium) & \cellcolor{gray!10}25,000 & \cellcolor{gray!10}\ding{52} \\
      Sound Event Detection & FSD50K & 4,009 & - \\
      Sound Event Detection & FSDKaggle2018 & 9,473 & - \\
      Genre Classification & GTZAN & 900 & - \\
      Speaker Counting & LibriCount & 4,576 & - \\
      Instruments Class & NSynth & 289,205 & - \\
      Keyword Spotting & SpeechCommand\_v1 & 51,088 & - \\
      Urban Sound Class. & UrbanSound8K & 7,895 & - \\
      Non-speech Sounds & VocalSound & 15,531 & - \\
      Binary Speaker ID & VoxCeleb1 & 22,576 & - \\
      Language ID & VoxLingua33 & 66,000 & - \\
      \midrule
      \multicolumn{4}{c}{\textit{Task 2 for Stage-1 and Stage-2 Training}} \\
      \midrule
      Speech Recognition & AISHELL-1 & 120,098 & - \\
      Sound Caption & Clotho & 3,839 & - \\
      Speech Recognition & LibriSpeech & 28,539 & - \\
      \multirow{3}{*}{Audio Caption} & MECAT & 20,039 & - \\
          & \cellcolor{gray!10}MACS & \cellcolor{gray!10}3,930 & \cellcolor{gray!10}\ding{52} \\
          & \cellcolor{gray!10}AudioSetCaps & \cellcolor{gray!10}1,729,996 & \cellcolor{gray!10}\ding{52} \\
      \multirow{3}{*}{Music Caption} & SongDescriber & 359 & - \\
          & \cellcolor{gray!10}MusicBench & \cellcolor{gray!10}52,768 & \cellcolor{gray!10}\ding{52} \\
          & \cellcolor{gray!10}MusicCaps & \cellcolor{gray!10}5,380 & \cellcolor{gray!10}\ding{52} \\
      % \midrule[0.3pt]\bottomrule[1pt]
      \bottomrule
    \end{tabular*}
    % \vspace{-0.4cm}
  \end{table}

In the first stage, we leverage large-scale sequence-level supervision from Automatic Speech Recognition (ASR) and audio captioning to map MoE-fused audio features into the LLM's embedding space, establishing preliminary audio--language alignment. We freeze both the audio encoders and the LLM while optimizing the MoE module and projector, thereby protecting the LLM's pre-trained linguistic priors from noise introduced by initially unaligned audio features. In the second stage, we keep the audio encoders frozen and jointly fine-tune the remaining components using diverse instruction-following data covering all 20 tasks. This allows the instruction $I$ to guide the extraction of task-relevant information from the aligned MoE representations. Despite employing a lightweight 135M-parameter LLM, the resulting model achieves strong cross-task generalization, highlighting the effectiveness of the fused audio representations and staged optimization.

\subsection{Task-Specific Data Scaling for Training}\label{sec:tsds}
% \noindent\textbf{Unbalaned Sample Analyses.} 
As shown in Table~\ref{tab:data_counts}, the training datasets vary substantially in scale, ranging from only a few hundred samples in SongDescriber to hundreds of thousands in NSynth. During unified multi-task optimization, this disparity can bias learning toward high-resource tasks, while low-resource tasks, particularly audio captioning, remain prone to overfitting and limited output diversity.

To mitigate this cross-task imbalance, we propose Task-Specific Data Scaling (TSDS), which selectively augments underrepresented tasks with task-relevant public datasets rather than uniformly scaling all tasks. The newly introduced datasets are highlighted in Table~\ref{tab:data_counts}. For captioning, we incorporate MusicBench~\cite{musicbench}, MusicCaps~\cite{musiclm}, MACS~\cite{macs}, and AudioSetCaps~\cite{audiosetcaps}, with AudioSetCaps alone providing approximately 1.73M audio--text pairs. These additions expand both acoustic and linguistic diversity, promoting robust cross-modal alignment and more diverse caption generation. For classification tasks affected by class imbalance, we supplement genre and emotion recognition with Free Music Archive~\cite{fma} and IEMOCAP~\cite{iemocap}, respectively, broadening category coverage and reducing bias toward dominant classes.

\section{Experimental Results and Analyses}

\subsection{Experimental Setups}
\noindent\textbf{Preprocess.} To adhere to the 24GB VRAM competition limit, we employ memory-efficient preprocessing: audio clips are uniformly truncated to a maximum duration of 30 seconds to prevent out-of-memory (OOM) errors, and individual samples are processed iteratively to further minimize peak memory usage during feature extraction. The resulting modality-specific inputs are then fed into the Qwen2-Audio and Audio-Flamingo 3 branches as Log-Mel spectrograms and raw waveforms, respectively.

\noindent\textbf{Training Data.} We train on diverse public datasets, including only the official training split of XARES-LLM and supplementary datasets for scaling. To improve multi-task learning and mitigate cross-task data imbalance, we apply Task-Specific Data Scaling (TSDS) tailored to each task. Details of TSDS and training data statistics are provided in Section~\ref{sec:tsds} and Table~\ref{tab:data_counts}, respectively.

\noindent\textbf{Training Details.} We train \tool for 30,000 and 80,000 steps in Stages 1 and 2, respectively, using three NVIDIA L20 GPUs. The MoE comprises four routed experts and one shared expert with top-2 routing ($K=2$), and each expert network has a hidden dimension of 3,413. As expert utilization remains balanced, we do not apply an auxiliary load-balancing loss.

\noindent\textbf{Evaluation.} We evaluate all models on the 20 public tasks of XARES-LLM~\cite{zhang2025xares} under a unified generative protocol that replaces task-specific heads with text prediction. Task~1 (public Track~A) comprises 15 speech, sound, and music classification datasets, evaluated by accuracy for single-label tasks and mean average precision (mAP) for multi-label tasks. Task~2 (public Track~B) comprises five ASR and audio-captioning datasets, evaluated by inverted character or word error rate (iCER/iWER) and FENSE or DATE, respectively. The overall XARES-LLM score is macro-averaged across all datasets.

The official Interspeech 2026 Audio Encoder Capability Challenge~\cite{dinkel2026audioencoderchallenge} further includes six hidden classification datasets and four hidden understanding datasets to assess generalization to unseen data. For each submission, the organizers freeze the encoder and train a lightweight projector and LoRA-adapted SmolLM2-135M decoder~\cite{allal2025smollm2}. Track~A and A~(Hidden) use separate decoders because of their different label spaces, whereas B~(Hidden) reuses the public Track~B decoder for out-of-domain evaluation.

\begin{table}[!ht]
    \centering
    \caption{Benchmarking results of various audio encoders}
    \label{tab:encoders}
    \small
    \renewcommand{\arraystretch}{1.05}
    \begin{tabular*}{\linewidth}{@{\extracolsep{\fill}}llrrr@{}}
      % \toprule[1pt]\midrule[0.3pt]
      \toprule
      \textbf{Encoder} & \textbf{Domain}
      & \textbf{Task 1} & \textbf{Task 2} & \textbf{Overall} \\
      \midrule
      \rowcolor{gray!10} \multicolumn{5}{c}{\textit{Audio Encoders for Specific Tasks.}} \\
      \midrule
      Whisper-large~\cite{whisper} & Speech & 0.698 & 0.441 & 0.634 \\
      FunASR-nano~\cite{funasr} & Speech & 0.552 & 0.516 & 0.543 \\
      MERT-330M~\cite{mert} & Music & 0.475 & 0.236 & 0.415 \\
      Dasheng-base~\cite{dasheng} & Unified & 0.540 & 0.233 & 0.463 \\
      Dasheng-1.2B~\cite{dasheng} & Unified & 0.587 & 0.236 & 0.499 \\
      \midrule
      \rowcolor{gray!10} \multicolumn{5}{c}{\textit{Audio Encoders from Large Audio Language Models.}} \\
      \midrule
      Kimi-Audio~\cite{kimi-audio} & Unified & 0.720 & 0.435 & 0.648 \\
      Qwen3-Omni~\cite{qwen3-omni} & Unified & 0.693 & 0.559 & 0.659 \\
      Qwen2.5-Omni~\cite{qwen2.5-omni} & Unified & 0.732 & 0.597 & 0.698 \\
      \textbf{Qwen2-Audio}~\cite{Qwen2} & Unified & 0.770 & \textbf{0.638} & 0.737 \\
      \textbf{Audio-Flamingo 3}~\cite{Flamingo} & Unified & \textbf{0.810} & 0.635 & \textbf{0.767} \\
      % \midrule[0.3pt]\bottomrule[1pt]
      \bottomrule
    \end{tabular*}
    % \vspace{-0.2cm}
  \end{table}

\subsection{Benchmarking Audio Encoder}\label{sec:encoders}
We evaluate representative encoders from speech (Whisper-large), music (MERT-330M), and general audio (Dasheng-1.2B) domains on the XARES-LLM benchmark (Table \ref{tab:encoders}). Notably, the audio encoders from pre-trained LALMs, specifically Qwen2-Audio and Audio-Flamingo 3, yield the highest average scores of 0.737 and 0.767, respectively (Table \ref{tab:encoders}). 

Our analysis suggests that single-encoder models often exhibit domain-specific biases or coverage deficiencies. Motivated by this, we propose \tool to leverage the complementary strengths of these leading LALM backbones for more robust unified audio modeling.

\subsection{MoE Makes Better Fusion}

We compare our MoE with three fusion baselines---Concatenation (Concat), Addition (Add), and Gating---across two encoder pairs (Table~\ref{tab:fusion}). Here, Gating denotes a dense encoder-level fusion strategy that combines both encoder representations using a learnable coefficient; unlike the sparse MoE router, it performs neither conditional expert selection nor token-level routing. Results show that Concat consistently outperforms Add and Gating. We attribute this to Concat's ability to preserve feature integrity and avoid the information blurring inherent in summation, thereby providing richer representations for the downstream LLM.

\begin{table}[!ht]
    \centering
    \caption{Comparison of model fusion strategies}
    \label{tab:fusion}
    \small
    \renewcommand{\arraystretch}{1.08}
    \begin{tabular*}{\linewidth}{@{\extracolsep{\fill}}lcccccc@{}}
      % \toprule[1pt]\midrule[0.3pt]
      \toprule
      & \multicolumn{3}{c}{(Q2A, AF3)}
      & \multicolumn{3}{c}{(Q2A, KA)} \\
      \cmidrule(r){2-4} \cmidrule(lr){5-7} 
      & Concat & Gating & \textbf{MoE (Ours)}
      & Concat & Add & \textbf{MoE (Ours)} \\
      \midrule
      Task 1 
          & 0.816 & 0.800 & \textbf{0.839} 
          & 0.785 & 0.778 & \textbf{0.816}  \\
      Task 2 
          & 0.662 & 0.660 & \textbf{0.691} 
          & 0.648 & 0.604 & \textbf{0.682}  \\
      Overall 
          & 0.778 & 0.764 & \textbf{0.802} 
          & 0.750 & 0.735 & \textbf{0.782} \\
      % \midrule[0.3pt]\bottomrule[1pt]
      \bottomrule
    \end{tabular*}
    % \vspace{-0.1cm}
  \end{table}

  \begin{table}[!ht]
    \centering
    \caption{Ablation study of the \tool}
    \label{tab:ablation}
    \small
    \renewcommand{\arraystretch}{1.08}
    \begin{tabular*}{\linewidth}{@{\extracolsep{\fill}}lrrrr@{}}
      % \toprule[1pt]\midrule[0.3pt]
      \toprule
      & Params (M) & Task 1 & Task 2 & Overall \\
      \midrule
      \textbf{\tool} 
          & 115.8  & \textbf{0.839} & 0.691 & \textbf{0.802} \\
      w/o SwiGLU 
          & 85.2 & 0.827 & \textbf{0.693} & 0.794 \\
      \tool-Light 
          & 88.5 & 0.832 & \textbf{0.693} & 0.797 \\
      % \midrule[0.3pt]\bottomrule[1pt]
      \bottomrule
    \end{tabular*}
    % \vspace{-0.1cm}
  \end{table}

Further exploration of encoder combinations (Table \ref{tab:main_results}) reveals that the dual-model setup (Qwen2-Audio and Audio-Flamingo 3) achieves the best baseline (0.778 on XARES-LLM). Interestingly, a triple-model configuration (adding Dasheng-1.2B) degrades performance, suggesting that indiscriminate encoder stacking introduces redundancy rather than complementarity. Leveraging the optimal dual-model backbone, our \tool further improves the score to 0.802. This significant gain over the best Concat baseline demonstrates that the MoE framework more effectively captures and exploits the synergies between complementary audio encoders.

\subsection{Ablation Study}

We first evaluate the impact of the SwiGLU activation function on the MoE experts. As shown in Table \ref{tab:ablation}, removing SwiGLU degrades the overall score from 0.802 to 0.794, confirming its importance in enhancing the non-linear expressivity of the MoE architecture.

To assess parameter efficiency, we introduce \tool-Light, a compressed variant with 88.5 M parameters. Despite its reduced footprint, \tool-Light maintains competitive performance (0.797), demonstrating the robustness of our architecture. These results validate the parameter efficiency of \tool and suggest that our data scaling strategy effectively compensates for reduced model capacity.

\subsection{Data Scaling Strategies}

To address task imbalance within XARES-LLM, we implement Task-Specific Data Scaling (TSDS) for music, emotion, and audio captioning. As shown in Table~\ref{tab:data_scale}, TSDS substantially improves the standard MoE from 0.759 to 0.794, whereas the higher-capacity MoE-SwiGLU exhibits a smaller aggregate gain from 0.799 to 0.802. This contrast suggests that the stronger architecture may already capture much of the available overall improvement. Nevertheless, TSDS remains important for low-resource or imbalanced tasks: on Free Music Archive, for example, it improves performance from 0.848 to 0.979, demonstrating that aggregate scores can understate its task-specific benefits.

\begin{table}[!ht]
    \centering
    \caption{The table compares data-scaling performance across tasks; ``w/o'' denotes training without TSDS, whereas ``w'' denotes training with TSDS.}
    \resizebox{\linewidth}{!}{
    \begin{tabular}{ccccc ccccc ccccc ccccccccccc}
      % \toprule[1pt]\midrule[0.3pt]
      \toprule
      \multirow{2}[2]{*}{Model}
      & \multirow{2}[2]{*}{Method}
      & \multicolumn{15}{c}{Task 1}
      & \multicolumn{5}{c}{Task 2} 
      & \multirow{2}[2]{*}{Overall}\\
      \cmidrule(l){3-17}\cmidrule(lr){18-22}
      & & \circled{1} & \circled{2} &\circled{3} 
      & \circled{4} & \circled{5} &\circled{6}
      & \circled{7} & \circled{8} &\circled{9}
      & \circled{10} & \circled{11} &\circled{12} &\circled{13}
      & \circled{14} & \circled{15} &\circled{16} &\circled{17}
      & \circled{18} & \circled{19} &\circled{20} & \\
      \midrule
      \multirow{2}[2]{*}{MoE} 
          & \textbf{Ours}
              & 0.987 & 0.856 & 0.905 & 0.994 & 0.932 & 0.253 & 0.858 & 0.929 & 0.207 & 0.771 & 0.966 & 0.886 & 0.947 & 0.973 & 0.939
              & 0.851 & 0.474 & 0.904 & 0.688 & 0.550 & \textbf{0.794}  \\
          & w/o TSDS
              & 0.981 & 0.841 & 0.890 & 0.993 & 0.923 & 0.204 & 0.808 & 0.909 & 0.184 & 0.765  & 0.928 & 0.885 & 0.944 & 0.976 & 0.920
              & 0.714 & 0.439 & 0.855 & 0.633 & 0.493 & 0.759  \\
          & w AudioSet
              & 0.992	& 0.797 & 0.875 & 0.994 & 0.87 & 0.247 & 0.875 & 0.909 & 0.191 & 0.747 & 0.964 & 0.886 & 0.941 & 0.978 & 0.943
              & 0.829 & 0.485 & 0.897 & 0.694 & 0.528 & 0.782\\
      \midrule
      \multirow{2}[2]{*}{\makecell[c]{MoE- \\SwiGLU}} 
          & \textbf{Ours}
              & 0.982 & 0.847 & 0.882 & 0.994 & 0.979 & 0.251 & 0.872 & 0.899 & 0.381 & 0.782 & 0.968 & 0.888 & 0.941 & 0.974 & 0.94
              & 0.882 & 0.449 & 0.914 & 0.7 & 0.512 & \textbf{0.802}\\
          & w/o TSDS
              & 0.993	& 0.847 & 0.9 & 0.995 & 0.848 & 0.254 & 0.87 & 0.9 & 0.375 & 0.79 & 0.974 & 0.889 & 0.941 & 0.978 & 0.938
              & 0.870 & 0.470 & 0.91 & 0.7 & 0.533 & 0.799 \\
          & w AudioSet
              & 0.984 & 0.818 & 0.858 & 0.995 & 0.836 & 0.248 & 0.875 & 0.909 & 0.267 & 0.791 & 0.972 & 0.873 & 0.942 & 0.978 & 0.937
              & 0.868 & 0.461 & 0.909 & 0.695 & 0.529 & 0.787\\
      % \midrule[0.3pt]\bottomrule[1pt]
      \bottomrule
    \end{tabular}
    }
    \label{tab:data_scale}
  \end{table}

  \begin{table}[!ht]
    \caption{Official leaderboard results of \tool in the Interspeech 2026 Audio Encoder Capability Challenge, covering Track A (classification) and Track B (audio understanding)}
  \label{tab:challenge_results}
  \centering
  \footnotesize
%   \scriptsize
  \setlength{\tabcolsep}{3pt}
  \renewcommand{\arraystretch}{1.02}
  \begin{tabular*}{\linewidth}{@{\extracolsep{\fill}}lrlrlr@{}}
    \toprule
    \multicolumn{2}{c}{\textbf{Public Tasks}} &
    \multicolumn{2}{c}{\textbf{Hidden Tasks}} &
    \multicolumn{2}{c}{\textbf{Aggregate Results}} \\
    \cmidrule(r){1-2}\cmidrule(lr){3-4}\cmidrule(l){5-6}
    \textbf{Dataset} & \textbf{Score} &
    \textbf{Dataset} & \textbf{Score} &
    \textbf{Result} & \textbf{Score} \\
    \midrule
    \rowcolor{gray!10} \multicolumn{6}{c}{\textit{Track A Performance}} \\
    \midrule
    ASVspoof 2015          & \textbf{0.9900} &
    Fingersnap             & 0.8750 &
    Overall                & \textbf{0.9122} \\
    CREMA-D                & \textbf{0.8793} &
    KeyScratch             & \textbf{0.9990} &
    Public test            & \textbf{0.8527} \\
    ESC-50                 & 0.9125 &
    King-ASR-457           & 0.9850 &
    Hidden test            & 0.9717 \\
    Fluent Speech Commands & 0.9939 &
    King-ASR-719           & 0.9810 & & \\
    Free Music Archive     & \textbf{0.9587} &
    King-ASR-876           & \textbf{1.0000} & & \\
    FSD50K                 & 0.3208 &
    King-ASR-955           & \textbf{0.9900} & & \\
    FSDKaggle2018          & 0.8844 & & & & \\
    GTZAN                  & 0.9091 & & & & \\
    LibriCount             & 0.4790 & & & & \\
    NSynth                 & \textbf{0.7881} & & & & \\
    Speech Commands V1     & \textbf{0.9352} & & & & \\
    UrbanSound8K           & 0.8889 & & & & \\
    VocalSound             & 0.9393 & & & & \\
    VoxCeleb1              & 0.9720 & & & & \\
    VoxLingua33            & 0.9391 & & & & \\
    \midrule
    \rowcolor{gray!10} \multicolumn{6}{c}{\textit{Track B Performance}} \\
    \midrule
    AISHELL-1              & \textbf{0.9094} &
    AISHELL-6              & 0.5830 &
    Overall                & 0.6299 \\
    Clotho                 & 0.4789 &
    LibriHeavy             & 0.7366 &
    Public test            & 0.7014 \\
    LibriSpeech            & 0.9247 &
    MusicCaps              & \textbf{0.5155} &
    Hidden test            & 0.5583 \\
    MECAT                  & \textbf{0.6966} &
    TACOS                  & 0.3983 & & \\
    SongDescriber          & 0.4973 & & & & \\
    \bottomrule
  \end{tabular*}
  \parbox{\linewidth}{\scriptsize\emph{Note.}
  All entries are official normalized leaderboard scores in $[0,1]$.
Task-specific raw metrics are linearly mapped to normalized scores by
the challenge organizers. The overall score for each track is the
arithmetic mean of its public- and hidden-test aggregate scores.
Boldface denotes the best result, including ties, among the 23
participating teams.}
\end{table}

For comparison, we also evaluate broad data scaling by incorporating AudioSet, AudioSetCaps, and ASR data during Stage~1. This strategy improves the standard MoE from 0.759 to 0.782 but provides no corresponding gain for MoE-SwiGLU, indicating that indiscriminately increasing data volume yields diminishing returns for high-capacity architectures. These results further motivate targeted rather than uniform data scaling.

\subsection{Official Challenge Evaluation}
\label{sec:challenge_results}
In addition to the controlled comparisons above, we report the official leaderboard results of \tool to the Interspeech 2026 Audio Encoder Capability Challenge.
Track A evaluates discriminative audio representations, whereas Track B focuses on audio understanding. Table~\ref{tab:challenge_results} presents both aggregate and dataset-level results for the two tracks.

On Track~A, \tool achieves top-tier overall performance, with consistently strong results on both the public and hidden tests. At the dataset level, it delivers top-tier performance on eight tasks, including five public tasks and three hidden tasks. These leading results span speech, music, environmental sound, and spoofing detection, demonstrating robust transfer across heterogeneous discriminative domains and highlighting the benefit of complementary expert fusion. Performance on FSD50K and LibriCount nevertheless indicates that multi-label sound-event recognition and speaker counting remain challenging.

On Track~B, \tool also achieves competitive overall performance and strong results on the public test. It delivers top-tier performance on the public AISHELL-1 and MECAT tasks and, notably, on the hidden MusicCaps task, demonstrating strong speech recognition and effective audio--language modeling across general audio and music. Nevertheless, its relatively lower hidden-test aggregate performance, particularly on AISHELL-6 and LibriHeavy, suggests that robust generalization under unseen generative conditions remains an important direction for future work.

\section{Conclusion}

This paper presents \tool, a unified audio encoder that integrates complementary encoders from Qwen2-Audio and Audio-Flamingo 3 through MoE-based fusion. We employ SwiGLU and shared experts to facilitate interaction between heterogeneous encoder representations, together with Task-Specific Data Scaling in a two-stage training framework to improve cross-task generalization. On the XARES-LLM benchmark, \tool achieves state-of-the-art performance with an overall score of \textbf{0.802}. In the official Interspeech 2026 Audio Encoder Capability Challenge, \tool achieves top-tier overall performance on Track~A, including top-tier results on eight datasets, while also demonstrating strong performance on Track~B. These results validate the effectiveness of sparse expert architectures for integrating heterogeneous audio encoders and establish a scalable approach to unified, general-purpose audio understanding.

{\small
\noindent\textbf{Acknowledgments.}
This work was supported by National Natural Science Foundation of China (62076144).
\par}

{\small
\noindent\textbf{Disclosure of Interests.}
The authors have no competing interests to declare that are relevant to the content of this article.
\par}

\normalem
\bibliographystyle{splncs04}
\bibliography{mybib}

\end{document}